\documentclass[twocolumn,showpacs]{revtex4}
\usepackage{epstopdf}
\usepackage{graphicx}
\usepackage{color}
\usepackage{dcolumn}
\usepackage{bm}

\def\gappeq{\mathrel{ \rlap{\raise.5ex\hbox{$>$}}
                      {\lower.5ex\hbox{$\sim$}} } }
\def\lappeq{\mathrel{ \rlap{\raise.5ex\hbox{$<$}}
                      {\lower.5ex\hbox{$\sim$}} } }

\begin{document}

\preprint{PRA}

\title{Synthetic magneto-hydrodynamics in Bose-Einstein condensates and routes to vortex nucleation}

\author{L. B. Taylor$^{1}$, R. M. W. van Bijnen$^{2}$, D. H. J. O'Dell$^3$, N. G. Parker$^{4}$,  S. J. J. M. F. Kokkelmans$^2$ and A. M. Martin$^1$}

\address{$^{1}$ School of Physics, University of Melbourne, Parkville,
Victoria 3010, Australia. \\ $^{2}$ Eindhoven University of
Technology, PO Box 513, 5600 MB Eindhoven, The Netherlands. \\$^{3}$
Department of Physics and Astronomy, McMaster University, Hamilton,
Ontario, L8S 4M1, Canada. \\$^4$ School of Food Science and
Nutrition, University of Leeds, LS2 9JT, United Kingdom.}

\date{\today}

\begin{abstract}
Engineering of synthetic magnetic flux in Bose-Einstein condensates [Lin {\it et al.}, Nature {\bf 462}, 628 (2009)] has prospects for attaining the high vortex densities necessary to emulate the fractional quantum Hall effect.    
We analytically establish the hydrodynamical behaviour of a condensate in a uniform synthetic magnetic field, including its density and velocity profile. Importantly, we find that the onset of vortex nucleation observed experimentally corresponds to a dynamical instability in the hydrodynamical solutions and reveal other routes to instability and anticipated vortex nucleation.
\end{abstract}

 \maketitle

One of the driving forces behind quantum degenerate gas research is the emulation of many-body condensed matter phenomenon \cite{Bloch08}.  For quantum degenerate Bose gases, achievements include the observation of Bloch oscillations \cite{Bloch} and the Mott Insulator superfluid transition \cite{Mott_Insulator_Superfluid}. Considerable attention has also been applied to the achievement of the fractional quantum Hall (FQH) regime in Bose-Einstein condensates (BECs) \cite{QHE}.  One of the drawbacks to using ultra-cold atoms to emulate  condensed matter phenomenon is the breaking of time reversal symmetry to emulate the effects of electromagnetic fields on charged particles. In the context of FQH physics considerable focus has been applied to the breaking of time reversal symmetry through rotation.

For BECs in rotating traps the nucleation of vortices  has been observed by several groups \cite{Rotation_Experiment}. Theoretical studies, in the Thomas-Fermi (TF) regime, have proved effective in calculating the rotation frequency at which vortices are nucleated \cite{Recati01,Sinha01,Parker06}. Such studies, which have been extended to dipolar BECs \cite{Dipolar}, agree with numerical predictions from the  Gross-Pittaevskii equation \cite{Parker06,Lundh03,Tsubota02,Kasamastsu03,Lobo04,Parker05,Parker06a,Wright08}.   However, to reach the FQH regime the number of vortices needs to be significantly larger than the number of bosons in the BEC. Typically, experiments are carried out in parabolic traps, defined by average in-plane trapping frequency $\omega_{\perp}$. When the rotation frequency, $\Omega_z$, is equal to $\omega_{\perp}$ the BEC becomes untrapped. Hence, the attainment of the FQH regime, which requires $\Omega_z \rightarrow \omega_{\perp}$, is an extremely challenging task. An alternative approach is to generate a synthetic vector potential, which breaks the time reversal symmetry of the problem, producing a synthetic magnetic field \cite{Dalibard10,Ruostekoski02,Jaksch03,Mueller04,Speilman09,Murray09,Lin09}.  
  
The recent work of Lin {\it et al.} \cite{Lin09} has demonstrated the effectiveness of such an approach, in a $^{87}$Rb BEC, through the experimental realization of a synthetic vector potential in the Landau gauge:  ${\bm A^*}=A_x^* {\hat x}$, corresponding to a synthetic magnetic field ${\bf B^*}=\nabla \times {\bf A^*}$. The authors showed that  vortices were nucleated at a critical synthetic magnetic field. Here we generalize the TF methodology used to successfully calculate the onset of vortex nucleation in rotating systems \cite{Sinha01,Parker06} to the case of synthetic magnetic fields in harmonically confined BECs. We analytically determine stationary solutions, including the aspect ratios, of a BEC and determine the critical synthetic magnetic field at which the stationary solutions become unstable. In the rotating case this instability corresponds to the onset of vortex nucleation. We find that this analysis predicts the synthetic magnetic field at which vortex nucleation occurred in the experiments of  Lin {\it et al.} \cite{Lin09}.  

 
Our starting point is a generalized form of the Gross-Pittaevskii equation, which provides a mean-field description of the condensate wave function, $\psi \equiv \psi({\bf r},t)$: 
\begin{eqnarray}
i \hbar \frac{\partial \psi}{\partial t}&=& \left\{\frac{\left[-i\hbar {\bf \nabla} - {q^*{\bf A^*}}\right]^2}{2m} + V({\bf r},t) + g\left|\psi \right|^2\right\} \psi, 
\label{GPE1}
\end{eqnarray}
where $m$ is the atomic mass, $V({\bf r},t)$ is the confining potential and  $g=4\pi \hbar^2 a_s/m$ defines the contact interactions, via the s-wave scattering length $a_s$.
 
Expressing $\psi=\sqrt{\rho} \exp\left[i \theta\right]$ in the TF limit we can rewrite the Gross-Pittaevskii equation as:
\begin{eqnarray}
\frac{\partial \rho}{\partial t}=-{\bf \nabla}\cdot\left[ \rho{\bm \nu} \right]
\label{Continuity}
\end{eqnarray}
\begin{eqnarray}
\frac{\partial {\bf v}}{\partial t}&=&-{\bf \nabla}\left[ \frac{\nu^2}{2}+ \frac{1}{m}\left(V+g\rho\right)\right].
\label{Velocity}
\end{eqnarray}
where the generalized velocity is ${\bm \nu}={\bf v}-q^*{\bf A^*}/m$, with ${\bf v}=\hbar/m \nabla \theta$. 

To investigate stationary solutions we  set $\partial \rho / \partial t = \partial {\bf v} / \partial t=0$ and solve for $\rho$ and ${\bf v}$.  Having found static solutions, they are not necessarily stable and so we analyze their dynamical stability. To do this we consider small perturbations ($\delta \rho$ and $\delta \theta$) to the stationary solutions. Then, by linearizing the hydrodynamical equations, the dynamics of such perturbations are described by \cite{Sinha01,Parker06}
 \begin{eqnarray}
 \label{stability}
\frac{\partial }{\partial t} \left[\begin{array}{c}
 \delta \theta \\
  \delta \rho \\
\end{array}
\right] = -\left[\begin{array}{cc}
 {\bm \nu} \cdot {\bf \nabla} & \frac{g}{\hbar} \\
\frac{\hbar}{m}{\bf \nabla} \cdot \rho_0
{\bf \nabla} & \left[\left({\bf \nabla} \cdot
{\bm \nu}\right)+{\bm \nu} \cdot {\bf \nabla} \right] \\
\end{array}
\right] \left[\begin{array}{c}
 \delta \theta \\
  \delta \rho \\
\end{array}
\right].
\label{Perturbations}
 \end{eqnarray}
 To investigate the stability of the BEC we look for eigenfunctions and eigenvalues of the above operator: dynamical instability arises when one or more eigenvalues possess a real positive part. The size of the real eigenvalues dictates the rate at which the instability grows. Imaginary eigenvalues relate to stable collective modes of the system, i.e., sloshing and breathing \cite{Bijnen10}. In order to find such eigenfunctions we consider a polynomial ansatz for the perturbations in the coordinates $x$, $y$ and $z$ of total degree $n$. For the examples considered below, all operators in Eq. (\ref{Perturbations}), acting on polynomials of degree $n$, result in polynomials of (at most) the same degree. Therefore,  Eq. (\ref{Perturbations}) can be rewritten as a scalar matrix operator, acting on vectors of polynomial coefficients, for which finding the eigenvectors and eigenvalues is a trivial task.

Below we consider the static solutions to Eqs. (2) and (3) for a BEC in a static elliptical harmonic trap
under the influence of a synthetic magnetic field in the $z$-direction. The harmonic trapping potential has the form  $V({\bf r},t)=m\omega_{\perp}^2\left(x^2\left(1-\epsilon\right)+y^2\left(1+\epsilon\right)+z^2\gamma^2\right)/2$.
In the $x-y$ plane the trap has a mean trap frequency $\omega_{\perp}$ and ellipticity $\epsilon$. The trap strength in the axial direction is specified by $\gamma=\omega_z/\omega_{\perp}$.
Since the synthetic magnetic field is defined in terms of the synthetic vector potential  there is more than one possible choice of  ${\bf A^*}$. We consider the two most popular choices, the Landau and symmetric gauges. In each case we find the following form of the generalized velocity provides exact stationary solutions to Eqs.~(2) and~(3): ${\bm \nu}=\alpha\omega_{\perp}(y{\bf {\hat i}}+x{\bf {\hat j}})-q^*{\bf A^*}/m$.
For the symmetric (S) and Landau (L) gauges the synthetic vector potential has the forms: ${\bf A^*}=-{\bf {\hat i}}B_z y/2 +{\bf {\hat j}}B_z x/2$ (S) and  ${\bf A^*}=-{\bf {\hat i}}B_z y$ (L),
such that ${\bf B^*}={\bf {\hat k}} B_z^*$.

Setting $\partial \rho/\partial t=\partial {\bf v} /\partial t=0 $ in Eqs.~(\ref{Continuity},\ref{Velocity}) the stationary solutions for the respective gauges are:
\begin{eqnarray}
0&=&\left(\alpha+\frac{{\tilde B}_z^*}{2}\right){\tilde \omega}_x^2+\left(\alpha-\frac{{\tilde B}_z^*}{2}\right){\tilde \omega}_y^2 \, \, \, \, \, \, \,  \, \, \, \, \, \, {\rm (S)} \label{alpha_S}\\
0&=&\left(\alpha+{\tilde B}_z^*\right){\tilde \omega}_x^2+\alpha{\tilde \omega}_y^2 \, \, \, \, \, \, \, \, \, \, \,\, \, \, \, \, \, \, \, \, \, \, \, \, \,  \, \,\, \, \, \, \, \, \, \, \, \, \, \, \, \, \, \, \,  {\rm (L)}
\label{alpha_L}
\end{eqnarray}
where the effective trap frequencies  are
\begin{eqnarray}
{\tilde \omega}_x^2&=&(1-\epsilon)+\alpha^2-\alpha {\tilde B_z^*} +\left(\frac{{\tilde B}_z^*}{2}\right)^2, \, \, \, \,   \, \, \, \, \, \, \, \, \,{\rm (S)}
\label{Omega_x_Symmetric} \\
{\tilde \omega}_x^2&=&(1-\epsilon)+\alpha^2, \, \, \, \, \, \, \, \, \, \, \, \, \, \, \, \, \, \, \,  \, \, \, \,  \, \, \, \, \, \, \, \, \, \, \, \, \, \, \, \, \, \, \, \, \,  \, \, \, \, \, \, \, \, \, \, \, \, \, \, \, \, \, {\rm (L)}
\label{Omega_x_Landau} \\
{\tilde \omega}_y^2&=&(1+\epsilon)+\alpha^2+\alpha {\tilde B}_z^* +\left(\frac{{\tilde B}_z^*}{2}\right)^2, \, \, \, \, \,  \, \, \, \, \, \, \,  \, {\rm (S)}
\label{Omega_y_Symmetric} \\
{\tilde \omega}_y^2&=&(1+\epsilon)+\alpha^2+2 \alpha {\tilde B}_z^* +\left({\tilde B}_z^*\right)^2, \, \,  \, \, \, \, \, \, \, \, \, \, \, \, {\rm (L)}
\label{Omega_y_Landau}
\end{eqnarray}
and  ${\tilde B}_z^*=q^*B_z^*/m\omega_{\perp}$. Eqs.~ (\ref{alpha_S},\ref{Omega_x_Symmetric},\ref{Omega_y_Symmetric}) [Eqs.~ (\ref{alpha_L},\ref{Omega_x_Landau},\ref{Omega_y_Landau})] can be solved analytically to determine $\alpha$ in the symmetric [Landau] gauge as a function of  $\epsilon$ and  ${\tilde B}_z^*$, with the solutions being independent of the axial trapping strength, $\gamma$ \cite{Recati01,Sinha01,Parker06}. 

In Fig.~1(a) we plot the solutions $\alpha$ for the symmetric and Landau gauges respectively, for $\epsilon=0$. In each case for ${\tilde B}_z^* < 2$ only one solution exists with two additional solutions, referred to as the upper and lower branch solutions, bifurcating at  ${\tilde B}_z^* =2$. Despite the quantitative differences in the solutions for $\alpha$ the generalized velocities, relevant in determining the physical properties of the BEC, are gauge invariant.  

The solutions for $\alpha$ [Fig.~1(a)] have qualitatively similar properties to the solutions obtained in the TF limit for a BEC in a rotating trap. This similarity is best observed through a comparison of the symmetric gauge solutions with solutions for a BEC in a rotating trap \cite{Recati01,Sinha01,Parker06}. In a rotating trap $\alpha$ is determined by Eq.  (\ref{alpha_S}), with ${\tilde B}_z^*/2$ replaced by the dimensionless rotation frequency about the $z$-axis (${\tilde \Omega}_z=\Omega_z/\omega_{\perp}$), with terms proportional to $({\tilde B}_z^*)^2$ in Eqs. (\ref{Omega_x_Symmetric},\ref{Omega_y_Symmetric}) set to zero. This difference  emphasizes the key advantage  a synthetic magnetic field has over physical rotation. For physical rotation, when ${\tilde \Omega}_z \rightarrow 1$, ${\tilde \omega}_{x}$ (${\tilde \omega}_y$) tends to zero for the upper (lower) branch solution, hence the BEC is no longer confined. In the synthetic magnetic field case, the term proportional to $({\tilde B}_z^*)^2$ prevents ${\tilde \omega}_{x,y}$ tending to zero and  the BEC remains confined. In principle, this can enable the attainment of high vortex densities and hence the emulation of  FQH physics.    

\begin{figure}
\centering
\includegraphics[width=8.5cm]{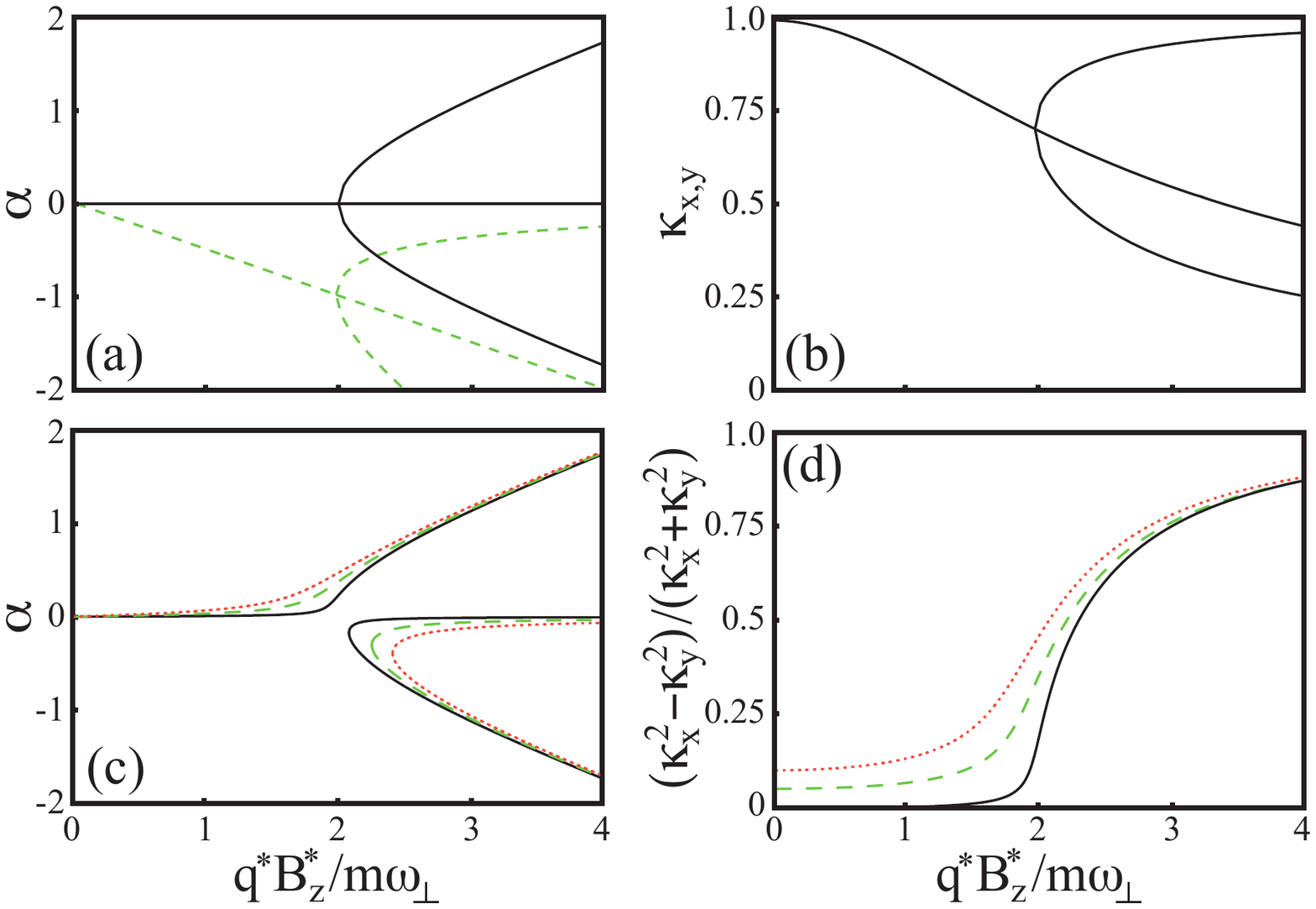}
\caption{(a) For a spherically symmetric trap, $\alpha$ as a function of the synthetic magnetic field strength, in the symmetric (solid black curve) and Landau (dashed green curve) gauges. (b) $\kappa_{x,y}$ as a function of ${\tilde B}_z^*$, in the symmetric and Landau gauges \cite{Note_Gauge}. (c) $\alpha$ as a function of the synthetic magnetic field (symmetric gauge). (d) the condensate ellipticity, $(\kappa_x^2-\kappa_y^2)/(\kappa_x^2+\kappa_y^2)$,  for the upper branch solutions in Fig.~1(c) as a function of the synthetic magnetic field strength \cite{Note_Gauge}. In (c)  and (d) $\epsilon=0.01$ (solid black curve), $\epsilon=0.05$ (dashed green curve) and $\epsilon=0.1$ (red dotted curve).}\label{Fig1}
 \vspace{-0.5cm}
\end{figure}

We now consider how the synthetic magnetic field alters the TF condensate density profile: 
\begin{eqnarray}
\rho=\rho_0\left(1-\frac{x^2}{R_x^2}-\frac{y^2}{R_y^2}-\frac{z^2}{R_z^2}\right),
\end{eqnarray} 
with $\rho_0=15N/(8\pi R_xR_yR_z)$. In Fig.~1(b)  we plot the aspect ratios of the condensate, $\kappa_{x,y}^2=R_{x,y}^2/R_z^2=\gamma^2/{\tilde \omega}_{x,y}^2$ [$R_z^2=2g\rho_0/(m\omega_z^2)$], 
as a function of the synthetic magnetic field, for both the symmetric and Landau gauges, with $\gamma=1$ and $\epsilon=0$. The solutions for $\kappa_{x,y}$ are gauge independent, with the solutions bifurcating at ${\tilde B}_z^*=2$. For low synthetic magnetic fields, ${\tilde B}_z^*<2$, the aspect ratios of the condensate in the $x$ and $y$ directions are reduced by $>$ 25\%, with $\kappa_x=\kappa_y$. For  ${\tilde B}_z^*>2$, consistent with the solutions obtained for $\alpha$ we have three solutions for $\kappa_{x,y}$. Since the solutions are gauge independent, with no loss of generality, we describe the dependence of the aspect ratios, for ${\tilde B}_z^*>2$, in terms of the solutions for $\alpha$ in the symmetric gauge [Fig. 1(a), solid black curve]. For the $\alpha=0$ solution  the aspect ratios of the condensate continue to reduce with $\kappa_x=\kappa_y$. For the upper (lower) branch solutions for $\alpha$  $\kappa_x$  follows the upper (lower) branch solutions  and $\kappa_y$  follows the lower (upper) branch solutions in Fig. 1(b), with $\kappa_x(\alpha)=\kappa_y(-\alpha)$.  

For $\epsilon \ne 0$ the solutions for $\alpha$ significantly change [Fig. 1(c)]. For the symmetric gauge, the $\alpha=0$ solution no longer exists and the plot has two distinct branches. The upper branch ($\alpha \ge 0$) is single valued and exists over all ${\tilde B}_z^*$. For this branch the ellipticity of the BEC in the $x-y$ plane  is plotted as a function of the synthetic magnetic field, in Fig.~1(d). We find that the BEC ellipticity monotonically increases from $\epsilon$ with increasing ${\tilde B}_z^*$. The lower branch ($\alpha <0$) is double valued and exists only when ${\tilde B}_z^*$ is greater than the backbending magnetic field: ${\tilde B}_{z {\rm (b)}}^*(\epsilon)$. This lower branch backbending point  shifts to higher ${\tilde B}_z^*$ as the ellipticity is increased  [solid green in Fig. 2(a) \cite{Note_Gauge}]. In Fig. 2(a) we have characterized the stability of the upper branch solutions through the evaluation of Eq.~(\ref{Perturbations}) for polynomial perturbations up to order $n=5$ \cite{Note_Gauge}. The instability region consists of a series of crescents \cite{Sinha01}. Each crescent corresponds to a single value of the polynomial degree $n$, where higher values of $n$ add extra crescents, from above. At the high synthetic field end these crescents merge to form  a main region of instability, characterized by large eigenvalues. At low synthetic magnetic  fields the crescents become vanishingly small, with eigenvalues several orders of magnitude smaller than in the main instability region.  These regions induce instability if they are traversed very slowly \cite{Corro07}. ${\tilde B}_{z {\rm (b)}}^*(\epsilon)$ and the stability of the stationary solutions of the upper branch are key to understanding the response of the BEC to the adiabatic introduction of $\epsilon$ or $\tilde{B}_z^*$.

For a fixed synthetic magnetic field as the ellipticity of the trap is introduced adiabatically, from zero, the BEC can follow two routes, depending on the value of ${\tilde B}_z^*$ relative to ${\tilde B}_{z {\rm (b)}}^*(0)$. For   ${\tilde B}_z^* < {\tilde B}_{z {\rm (b)}}^*(0)$  the solutions follow the upper branch  until these solutions become dynamically unstable. This route to instability is schematically indicated by the vertical black arrow in Fig. 2(a).  For  ${\tilde B}_z^* > {\tilde B}_{z {\rm (bif)}}^*(0)$ the BEC follows the lower branch  from $\alpha=0$ to negative $\alpha$. However, as $\epsilon$ is increased ${\tilde B}_{z {\rm (b)}}^*(\epsilon)$ shifts upwards [solid green curve in Fig 2(a)]. When ${\tilde B}_z^* > {\tilde B}_{z {\rm (b)}}^*(\epsilon)$ the lower branch no longer exists. This route to instability is schematically indicated by the vertical white arrow in Fig. 2(a). For fixed ellipticity as the synthetic magnetic field is introduced adiabatically, from zero, the BEC follows the upper branch solutions for $\alpha$, which become dynamically unstable. This route to instability is schematically indicated by the horizontal black arrow in Fig. 2(a)  

In the case of rotation \cite{Parker06} the analogous instabilities due to: (i) the dynamical instability of the upper branch solutions or (ii) the shifting of the backbending point lead to the nucleation of vortices. As such we expect that for: (i) the adiabatic introduction of a synthetic magnetic field, for a fixed trap ellipticity or (ii) the adiabatic introduction of trap ellipticity, for a fixed synthetic magnetic field, the onset of vortex nucleation can be determined by the onset of instability in the upper branch solutions and the evolution of the backbending point.

\begin{figure}
\centering
\includegraphics[width=8.5cm]{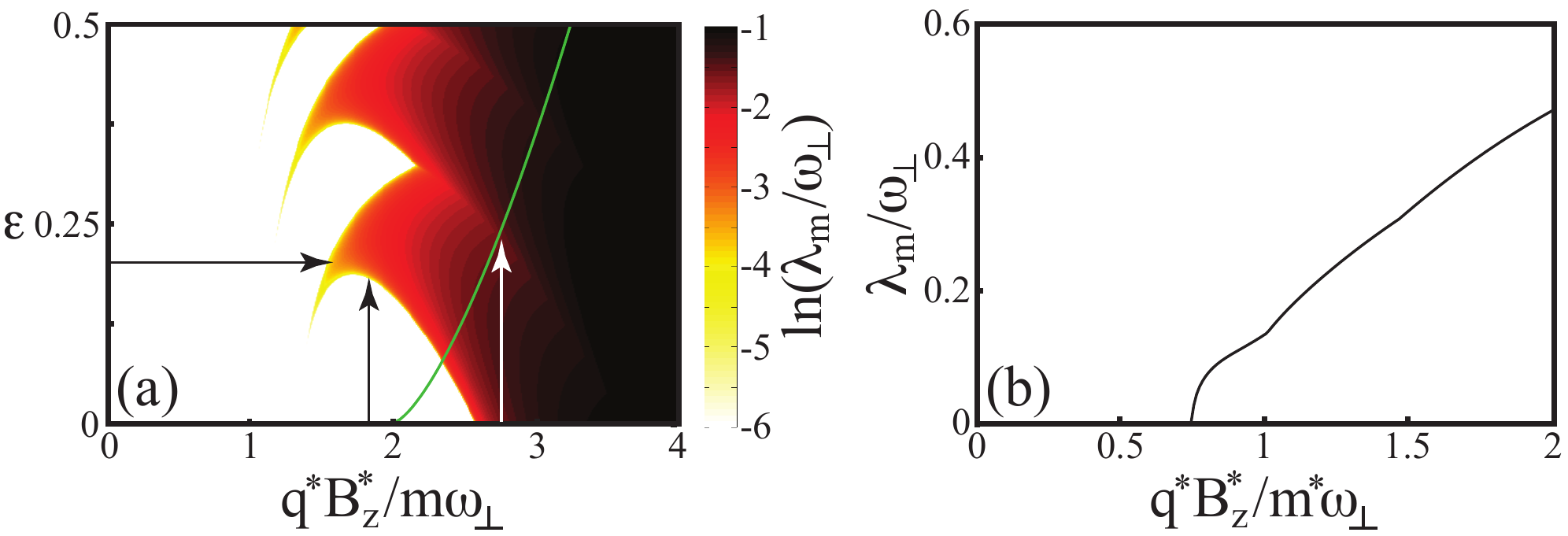}
\caption{(a) The logarithm of the maximum positive real eigenvalues of Eq.~(\ref{Perturbations}) [$n=5$] for the upper branch solutions of $\alpha$ as a function of ${\tilde B}_z^*$ and $\epsilon$. The solid green curve plots the trap ellipticity ($\epsilon$) at which the backbending point (${\tilde B}_{z({\rm b})}^*$) occurs \cite{Note_Gauge} and the horizontal/vertical arrows indicate routes to instability. (b) For the experimental parameters of Lin {\it et al.} \cite{Lin09,Expt_Param}, for the $\alpha$ branch which starts at $\alpha=0$ for ${\tilde B}_z^*=0$, the maximum positive real eigenvalues of Eq.~(\ref{Perturbations}) [$n=5$].}\label{Fig2}
 \vspace{-0.5cm}
\end{figure}

To test this we consider the recent experiment  of Lin {\it et al.} \cite{Lin09} where the synthetic magnetic field was generated via an engineered, Landau gauge, vector potential \cite{Speilman09}. To engineer the vector potential an $^{87}$Rb BEC was illuminated with a pair of Raman laser beams with a momentum difference along the $x$-direction, producing three tunable dressed states. The dressed state with the lowest energy corresponds to the following Hamiltonian along the $x$-direction $H_x=(\hbar k_x-q^*A_x^*)^2/2m^*$. In this Hamiltonian $A_x^*$ is the synthetic vector potential, controlled by the Zeeman shift for the atoms with synthetic charge $q^*$ and an effective mass for the atoms along the $x$-direction $m^*$.  To produce the desired spatially synthetic vector potential the Zeeman shift varies linearly in the $y$-direction, achieved by the application of a magnetic field gradient along $y$, resulting in an effective vector potential: $A_x^*=-B_z^*y$.  Including the effective mass into the hydrodynamical analysis: 
\begin{eqnarray}
 {\bm \nu}&=&\left(\frac{\hbar}{m^*}\frac{\partial \theta}{\partial x}+\frac{q^* B_z^*y}{m^*}\right){\bf {\hat{i}}}+\frac{\hbar}{m}\frac{\partial \theta}{\partial y}{\bf {\hat{j}}}\nonumber \\
 &=&\left(\frac{m}{m^*}\alpha \omega_{\perp} y+\frac{q^* B_z^*y}{m^*}\right){\bf {\hat{i}}}+\alpha \omega_{\perp} x{\bf {\hat{j}}},
\end{eqnarray}
the stationary solutions are determined by 
\begin{eqnarray}
0=\left(\frac{m}{m^*}\alpha+{\tilde B}_z^*\right){\tilde \omega}_x^2+\alpha{\tilde \omega}_y^2
\end{eqnarray}
with  ${\tilde B}_z^*=q^*B_z^*/(m^*\omega_{\perp})$, ${\tilde \omega}_x^2$ defined by Eq.~(\ref{Omega_x_Landau}) and 
\begin{eqnarray}
{\tilde \omega}_y^2=(1+\epsilon)+\frac{m}{m^*}\alpha^2+ \alpha {\tilde B}_z^* + \frac{m^*}{m}\left({\tilde B}_z^*\right)^2 .
\end{eqnarray} 


In the experiments the synthetic magnetic field is ramped up over 0.3s. Thus as the synthetic field is increased the solution starting at  $\alpha=0$ when ${\tilde B}_z^*=0$ will be tracked, until it becomes dynamically unstable. 
 Experimentally  vortex nucleation started when the synthetic magnetic flux ($\Phi_{B_z^*}=\pi R_x R_y B_z^*$), in units of the flux quantum ($\Phi_0=h/q^*$), passing through the BEC
\begin{eqnarray}
\frac{\Phi_{B_z^*}}{\Phi_0} =\frac{{\tilde B}_z^*}{2}\left(15 N \tilde{a}_s \gamma\right)^{\frac{2}{5}} \left({\tilde \omega}_x{\tilde \omega}_y\right)^{-\frac{3}{5}}\approx10.
\label{Expt}
\end{eqnarray}
In Eq.~(\ref{Expt}) we have redefined the contact interactions to include the effective mass difference in the $x$-direction, such that $ \tilde{a}_s=a_s/(2.5^{1/3}l_{xy})$, with $l_{xy}=\sqrt{\hbar/m\omega_{\perp}}$.  
Using the experimental parameters  \cite{Expt_Param} the stationary solution becomes unstable when ${\tilde B}_z^*=0.74$ [Fig. 2(b)], corresponding to $\frac{\Phi_B}{\Phi_0}=11.1$, in close agreement with the experimental observation of the onset of vortex nucleation. This strongly suggests that the observed onset of vortex nucleation arises from a dynamical instability as calculated from the stationary TF solutions  \cite{Nick_Note}.     

 The engineering of synthetic magnetic fields, in BECs, offers a new and exciting route into the emulation of FQH physics. The experimental advantages of such an approach, as compared to rotating systems, is the ability to enter the FQH regime with the number of vortices exceeding the number of bosons. Such an arrangement also provides a new route to investigate the FQH regime in lattice systems, without the need to rotate the lattice. We have shown that the nucleation of vortices, the first step in obtaining high density vortex systems, due to adiabatic changes in the synthetic field and ellipticity of the trap can be calculated in the TF regime. This enables us to analytically determine the aspect ratios of the BEC for fields below the nucleation point and show that the experimental observation of the onset of vortex nucleation  is due to dynamical instabilities in the stationary state of the BEC. 

We acknowledge financial support from the Natural
Sciences and Engineering Research Council of Canada (DHJOD) and the TU/e Fund for Excellence (RMWvB and SJJMFK).


\begin{thebibliography}{99}
\bibitem{Note_Gauge} This result is independent of the chosen gauge.
\bibitem{Nick_Note} Experimental parameters, $N{\tilde a}_s=350$, imply the system is in the TF regime.
\end{thebibliography}
\end{document}